\documentclass[%
 reprint,
 amsmath,amssymb,
 aps,
]{revtex4-2}

\usepackage{graphicx}
\usepackage{dcolumn}
\usepackage{bm}
\usepackage{hyperref}
\usepackage{url}
\usepackage{physics}
\usepackage{multirow}
\usepackage{braket}
\usepackage{float}
\usepackage{xcolor}

\begin{document}


\title{Supernovae neutrino detection via coherent scattering off silicon nuclei}

\author{A. L. Foguel}
\altaffiliation[Current address: ]{Departamento de F\'isica Matem\'atica, Instituto de F\'isica, Universidade de S\~ao Paulo, C.P. 66.318, S\~ao Paulo, 05315-970, Brazil}
\author{E. S. Fraga}%
 
 \author{C. Bonifazi}%
 
\affiliation{%
Instituto de F\'isica, Universidade Federal do Rio de Janeiro,
Caixa Postal 68528, 21941-972, Rio de Janeiro, RJ, Brazil}%

\date{\today}

\begin{abstract}
Low-energy neutrinos are clean messengers from supernovae explosions and probably carry unique insights into the process of stellar evolution. We estimate the expected number of events considering coherent elastic scattering of neutrinos off silicon nuclei, as would happen in Charge Coupled Devices (CCD) detectors. The number of expected  events, integrated over a window of about $18$ s, is $\sim 4$ if we assume $10$ kg of silicon and a supernovae $1$ kpc away. For a distance similar to the red supergiant Betelgeuse, the number of expected events increases to $\sim 30-120$, depending on the supernovae model. We argue that silicon detectors can be effective for supernovae neutrinos, and might possibly distinguish between models for certain target masses and distances.
\end{abstract}

\maketitle


\section{\label{sec:level1}Introduction}

Low-energy neutrinos play an important role in supernovae (SN) explosions and are responsible for carrying away most of the energy of the collapse \cite{Scholberg:2012id,Mirizzi:2015eza,Janka:2016fox,Janka:2017vlw}. It also affects mechanisms of cooling of proto-neutron stars, and their future detection might bring insight into the process of formation and evolution of compact stars \cite{Glendenning:1997wn,Horiuchi:2017sku}. Moreover, (anti-)neutrino signals could, in principle, also carry unique information on the possible phase transition from hadronic to quark matter if it occurs in the early post-bounce phase of a core-collapse supernovae \cite{Gentile:1993ma,Drago:1997tn,Nakazato:2008su,Sagert:2008ka,Dasgupta:2009yj,Buballa_2014}.

So far, however, there has been only one detection of supernovae neutrinos, in the event of Supernova SN1987a. At that time, the detection capabilities of the available detectors were orders of magnitude inferior than today, and e.g. the time sequence of the events was, unfortunately, statistically not significant \cite{Hirata:1987hu,Hirata:1988ad}, thereby precluding more detailed information on the phases of the proto-neutron star. 

Through the course of its life, previous to its final stages, the hydrostatic equilibrium of a star is maintained by the counterbalance of two opposite forces, gravity and the force generated from thermal pressure of fuel burning \cite{Shapiro:1983du}. If the star is massive enough, the burning proceeds to heavier elements, forming an onion shell structure inside the star. For stars heavier than $\sim 8 \; M_{\odot}$, the gravitational force is strong enough to trigger neutronization, i.e. the process in which a proton captures an electron releasing a neutron and an electron neutrino via the inverse beta decay
\begin{equation}\label{eq:neutronization}
    p + e^- \rightarrow n + \nu_e \; .
\end{equation}

Since neutrinos interact very weakly with matter, they quickly escape from the star, creating a pressure gradient that leads to the instability and collapse with a potentially supernova explosion.

The neutrinos emitted from the supernova carry away almost $99\%$ of the star binding energy. Their energy range lies between $0$ and $50$ MeV, the same energy for which coherent elastic scattering is expected to be dominant. Therefore, one could, in principle, detect supernova neutrinos using coherent elastic neutrino-nucleus scattering. A pioneering study behind the idea of neutrino detection via this channel is Ref. \cite{PhysRevD.30.2295}, where they considered several $\nu$ sources and target materials.

Furthermore, supernovae are the most powerful sources of MeV neutrinos in the Universe, which come in all flavors and are emitted over a timescale of several tens of seconds. Therefore, since the coherent elastic neutrino-nucleus scattering is a neutral current process, it can be used for detecting all three neutrino and antineutrino flavors.

In this paper, we estimate the expected number of supernovae neutrino (SN$\nu$) events considering a coherent elastic neutrino-nucleus scattering (CE$\nu$NS) between a neutrino and a silicon nucleus. Silicon has been gaining prominence in particle physics due to the increasing use of CCDs (Charge Coupled Devices) in experiments such as CONNIE (Coherent Neutrino Nucleus Interaction Experiment) \cite{Aguilar-Arevalo:2019jlr, Aguilar_Arevalo_2016}, SENSEI (Sub-Electron-Noise Skipper-CCD Experimental Instrument) \cite{Abramoff_2019, Crisler_2018} and DAMIC-M (Dark Matter in CCDs at Modane) \cite{Castello-Mor:2020jhd}. The calculated number of expected SN$\nu$ events, integrated over a window of about $18$ s, is in the order of $4$ events considering $10$ kg of silicon mass and a SN located $1$ kpc away. For a SN at a distance similar to the red supergiant Betelgeuse ($196$ pc away), and considering $10$ kg of detector mass, the number of expected events increases to $\sim 30-120$ depending on the SN model. Such results confirm that silicon detectors can be effectively employed as SN$\nu$ detectors and even distinguish between SN models for certain target masses and SN distances.

The paper is organized as follows. In Section \ref{sec-theory} we summarize the well-know results for the cross section for the coherent elastic neutrino-nucleus scattering within the Standard Model to lowest order. In Section \ref{SN-model} we present the SN model simulations we used to extract the neutrino emission spectra. In Section \ref{event-rate} we present and discuss our results for the neutrino event rate that would be obtained in a silicon detector under the conditions mentioned above. Finally, in Section \ref{outlook} we present our outlook and final comments.

\section{Coherent elastic neutrino-nucleus scattering cross section}
\label{sec-theory}

Predicted for the first time in 1974 by D. Z. Freedman \cite{FreedmanPhysRevD}, and recently measured by the COHERENT collaboration \cite{Akimov:2017ade,collaboration2020detection}, the coherent elastic neutrino-nucleus scattering is a neutral current interaction in which a neutrino of any flavor scatters off a nucleus transferring some energy in the form of nuclear recoil. 

The cross section of this process, considering a nucleus at rest with spin zero, $Z$ protons and $N$ neutrons, and neglecting radiative corrections, is well defined within the standard model and can be written as
\begin{eqnarray}\label{eq:crossSection}
    \dv{\sigma}{E_{nr}} = \frac{G_F^2}{8\pi} Q_W^2 \left[ \vphantom{\frac{2E_{nr}}{E_{\nu}}} \right. 2 &-& \left. \frac{2E_{nr}}{E_{\nu}} + \left(\frac{E_{nr}}{E_{\nu}}\right)^2
    \right. \nonumber \\
    &-& \left. \frac{M E_{nr}}{E_{\nu}^2} \right] M |F(q)|^2 \;,
\end{eqnarray}
where $G_F$ is the Fermi coupling constant, $E_{nr}$ is the nuclear recoil energy, $E_{\nu}$ is the neutrino energy, $M$ is the target mass, $F(q)$ is the nuclear form factor and
\begin{equation}
    Q_W = N - (1-4 \sin^2 \theta_W)Z
\end{equation}
is the weak nuclear charge, with $\theta_W$ being the weak mixing angle. 
For the typical (low) energies of CE$\nu$NS, $sin^2 \theta_W \approx 0.23$ \cite{Olive_2014}.

\begin{figure}
\includegraphics[scale= 0.5]{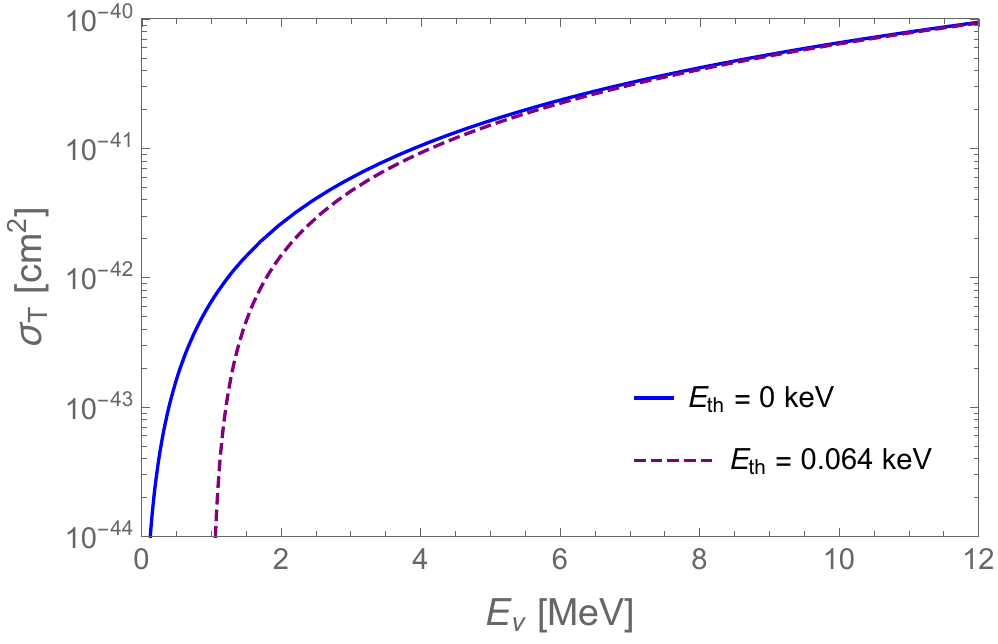}
\caption{\label{fig:cross_section} Total cross section of the coherent interaction between a neutrino and a silicon nucleus.}
\end{figure}

The nuclear form factor can be taken from Ref. \cite{PhysRevLett.84.2330}, yielding
\begin{equation}
    F(q) = \frac{4 \pi \rho_0}{A q^3} \left(\sin qR - qR \cos qR \right)\frac{1}{a+(aq)^2} \;,
\end{equation}
where $A$ is the atomic mass of the nucleus, $a = 0.7$ fm is the range of the considered Yukawa potential, $R = r_0 A^{1/3}$ is the nuclear radius and $\rho_0 = 3/(4\pi r_0^3)$ is the nuclear density. The average radius of a proton in a nucleus, $r_0$, is equal to $1.3$ fm. Considering that the coherent scattering occurs when the momentum transfer between the neutrino and the nucleus is sufficiently small, so that $q^2R^2<1$, one can show that $F(q) \approx 1$ in this limit. This condition means that the incident neutrino energy must lie in the MeV range, typically between 0 and 50 MeV.

Figure \ref{fig:cross_section} shows the total cross section $\sigma_T$ of the coherent scattering of neutrinos off silicon nuclei as a function of the neutrino energy. $\sigma_T$ is calculated by integrating Eq. (\ref{eq:crossSection}) over all possible values of the nuclear recoil energy starting from the energy threshold $E_{th}$ of the detector. In the figure, we illustrate the behavior of the total cross section by showing results for two cases: no threshold and the threshold from the CONNIE experiment (see below for further discussion).

\section{Supernova Model}
\label{SN-model}

In order to estimate the expected number of SN$\nu$ events that would be observed using coherent elastic scattering in CCDs, one of the most important ingredients is the SN neutrino spectrum. This information can only be obtained through SN numerical simulations. In this work, we focus on the results from two different SN simulations. 

The first one, by Nakazato \textit{et al.} \cite{Nakazato_2013}, uses one-dimensional simulations of neutrino-radiation hydrodynamics ($\nu$RHD) for the early phase of the collapse and quasi-static evolutionary calculations of proto-neutron star cooling (PNSC) with neutrino diffusion for the late phase.  The neutrino energy-spectra integrated from the core collapse to about $20$ s is available in a public database \cite{Nakazato_Database}. The light curves and spectra of SN neutrino were calculated considering three different parameters: initial mass of the progenitor star $M_p$, galaxy metallicities $Z$ and shock revival times $t_{rev}$ (for details, see Ref. \cite{Nakazato_2013}). 

\begin{figure}
\vspace{2mm}
\includegraphics[scale= 0.45]{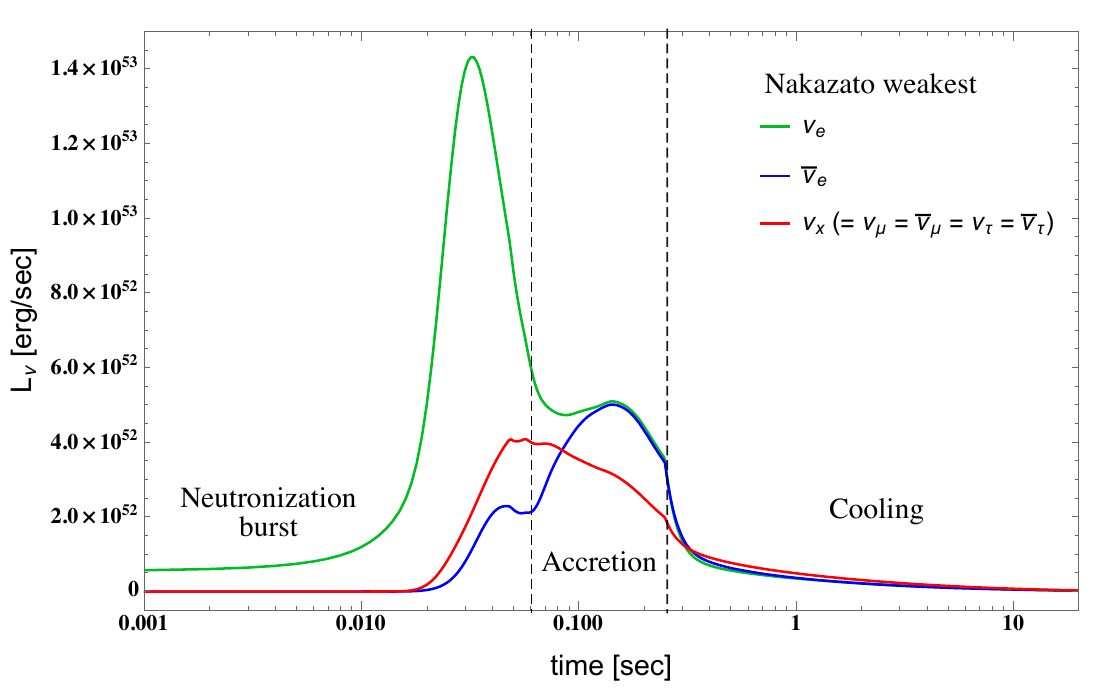}
\caption{\label{fig:luminosity} Time profile of the neutrino luminosity considering the Nakazato simulations with $M_p = 20 \;M_{\odot}$, $Z=0.02$ and $t_{rev}=200$ ms. The green peak signals the neutronization burst of electron neutrinos.}
\end{figure}

In what follows, we consider three different Nakazato scenarios: Nakazato weakest ($M_p = 20 \;M_{\odot}$, $Z=0.02$, $t_{rev}=200$ ms), Nakazato brightest ($M_p = 30 \;M_{\odot}$, $Z=0.02$, $t_{rev}=300$ ms) and Nakazato black hole ($M_p = 30 \;M_{\odot}$, $Z=0.004$). The nomenclature and specific configurations that were adopted here are the same ones used in the paper by the XMASS Collaboration \cite{Abe_2017}, where the authors calculated the number of supernova neutrino events that would be recorded in the XMASS experiment through the coherent scattering on xenon nuclei.

Figure \ref{fig:luminosity} shows the time evolution of the neutrino luminosity during the collapse for the Nakazato weakest model. The peak of electron neutrinos that appears just after the bounce ($t=0$) marks the neutronization burst. This burst is caused by the propagation of the shock wave through the core, which dissociates nuclei into free nucleons that quickly undergo the electron capture reaction, Eq. (\ref{eq:neutronization}), producing a huge quantity of electron neutrinos.

The $\nu_e$ burst is followed by an accretion phase, where all flavors of neutrinos are produced via thermal ($\nu_x$) and charged current ($\nu_e$ and $\bar \nu_e$) processes. In this stage $\nu_e$ and $\bar \nu_e$ luminosities are dominant due to their large emission by electron and positron capture.
 
However, the mass accretion eventually stops and the proto-neutron star enters the Kelvin-Helmholtz cooling phase. Now, the only source of neutrino emission is by diffusive transport from the dense and hot core. 
Hence, the luminosities from the different neutrino flavors become very similar, and gradually decrease while the star releases its remaining gravitational binding energy (for reviews on supernova neutrino emission, see \cite{Horiuchi:2017sku, Mirizzi:2015eza, Muller:2019upo}).

\begin{figure}
\includegraphics[scale= 0.43]{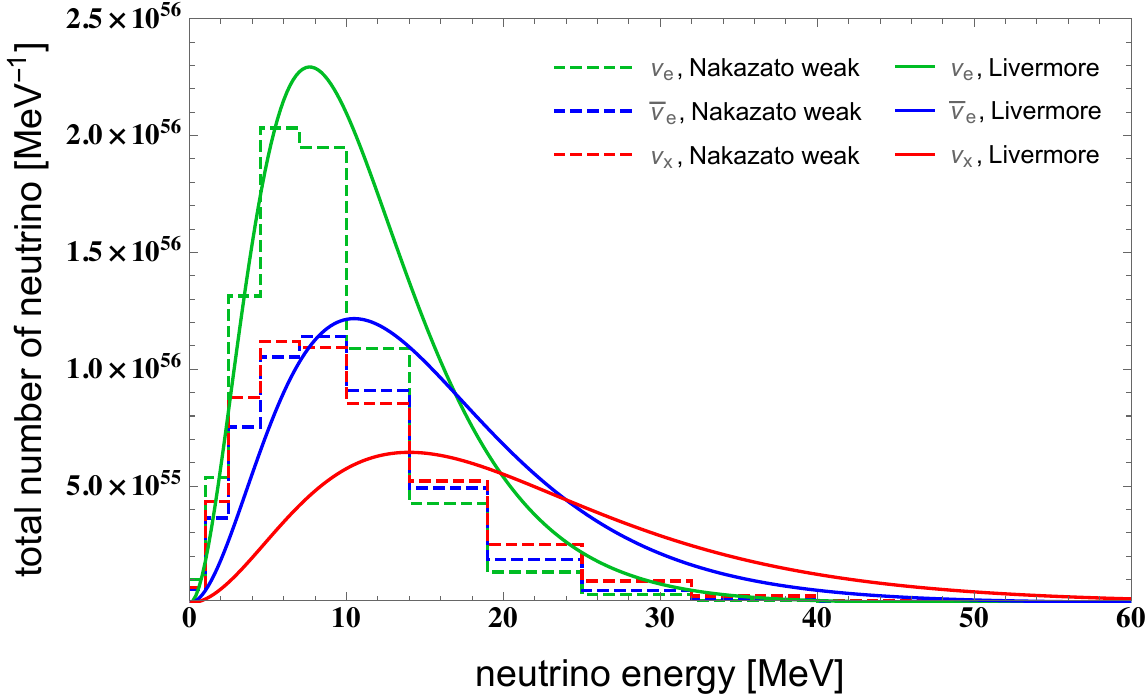}
\caption{\label{fig:spectra} Total number of emitted neutrinos, integrated to about $18$ s after the core bounce, as a function of the neutrino energy. The colors indicate different neutrino flavors, where $\nu_x = (\nu_{\mu}+ \bar \nu_{\mu}+ \nu_{\tau} + \bar \nu_{\tau})/4$. Solid lines represents the Livermore SN model \cite{Totani_1998} and dashed lines the Nakazato weakest model \cite{Nakazato_2013}.}
\end{figure}

The second SN simulation considered in our study is the Livermore model from Totani \textit{et al.} \cite{Totani_1998}. This model is a one-dimensional numerical simulation based on SN1987A and performed from the onset of the collapse to $18$ s after the core bounce. The progenitor is a main-sequence star of about $20 M_{\odot}$. Figure \ref{fig:spectra} shows the comparison of the supernova neutrino energy spectra from both models, Livermore and Nakazato weakest. From the figure, one can see that the Livermore model predicts a greater emission of higher energy neutrinos. A common feature of all simulations is that the neutrino mean energies obey the same hierarchy $\braket{E_{\nu_e}}<\braket{E_{\bar \nu_e}}<\braket{E_{\nu_x}}$ over the course of the supernova stages, where $\nu_x$ represents other flavor contributions, i.e. $\nu_x = (\nu_{\mu}+ \bar \nu_{\mu}+ \nu_{\tau} + \bar \nu_{\tau})/4 $. For references on other SN$\nu$ simulations, see \cite{Scholberg:2012id, Nagakura:2020qhb}.

%

\section{Event Rate and Discussion}
\label{event-rate}

The differential event rate of supernova neutrinos as a function of the nuclear recoil energy $E_{nr}$ can be calculated from the differential cross section  $(d\sigma / dE_{nr})$ of the coherent scattering as (see e.g. \cite{Abe_2017}):
\begin{equation}\label{eq:diffEventRate}
    \frac{\text{d} R_0}{\text{d} E_{nr}} (E_{nr}) = \frac{M_{det}N_A}{A(4\pi d^2)} 
    \sum_{i=\nu_e,\bar \nu_e, \nu_x} \int_{E_{min}}^{\infty} \dv{\sigma}{E_{nr}}(E_{\nu},E_{nr})f_i(E_{\nu}) \mathrm{d}E_{\nu} \; ,
\end{equation}
where $M_{det}$ is the detector mass, $d$ is the supernova distance, $N_A$ is  Avogadro’s number, $A$ is the averaged atomic mass of silicon, $f_i(E_{\nu})$ is the neutrino energy spectrum and $E_{min} = (E_{nr}+ \sqrt{E_{nr}^2 + 2ME_{nr}})/2$ is the minimal energy that the SN neutrino must have to produce a nuclear recoil with energy $E_{nr}$. 

Due to the fact that the energy deposition in CE$\nu$NS is very low (nuclear recoil energies are in the keV range \cite{Scholberg_2015}), it is experimentally very difficult to detect this interaction, being a measurement that requires very sensitive and low-noise detectors. One possibility are Charge Coupled Devices (CCDs), silicon sensors with low energy threshold that have been recently gaining prominence in neutrino and dark matter direct-detection experiments \cite{Crisler_2018, Abramoff_2019, Aguilar_Arevalo_2016_DAMIC, Castello-Mor:2020jhd}. 

\begin{figure}
\vspace{3mm}
\includegraphics[scale= 0.51]{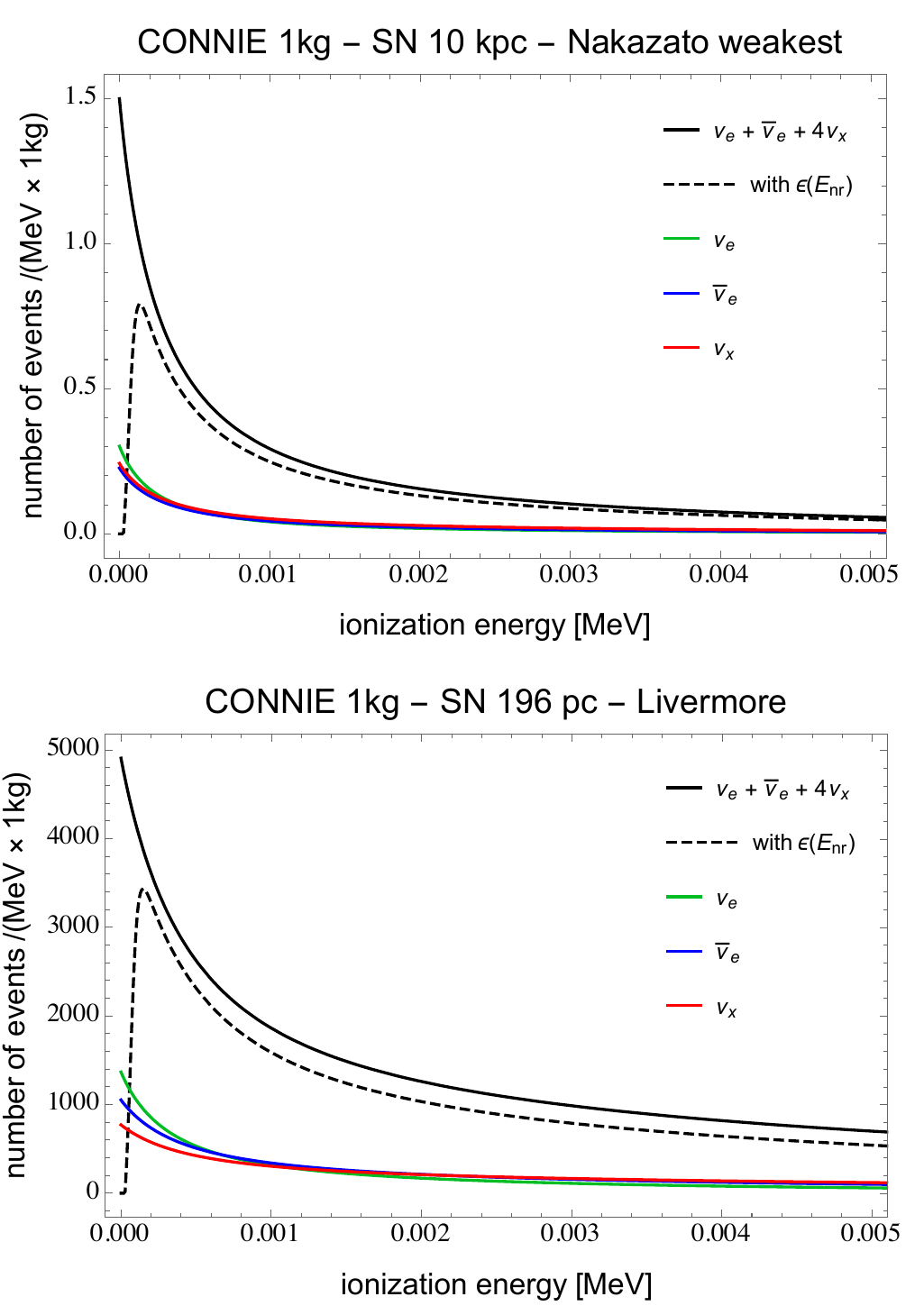}
\caption{\label{fig:eventRateNakWeak}
Energy spectrum as a function of the silicon ionization energy calculated from Eq. (\ref{eq:diffEventRatequenching}) for the Nakazato weakest SN model (upper plot) and the Livermore SN model (lower plot). The SN distance was set to $10$ kpc in the upper panel and to $196$ pc in the lower panel. In both plots 
the detector mass is equal to $1$ kg. The solid black curve shows the contribution of all neutrino and antineutrino flavors. The dashed black curve is the same as the solid but considering the processing efficiency of CONNIE, according to Eq. (\ref{eq:diffEventRateEff}). The green, blue and red lines show the differential event rate, but considering only $\nu_e$, $\bar \nu_e$ and $\nu_x (=\nu_{\mu} = \bar \nu_{\mu} = \nu_{\tau} = \bar \nu_{\tau})$, respectively.}
\end{figure}

Let us consider one particularly illustrative example of experiment that employs CCDs as  solid-state detectors, so that we can extract typical values for some observables: the CONNIE experiment. Located $30$ m from the core of the Angra 2 nuclear power plant in the state of Rio de Janeiro, Brazil, the experiment has as its main goal the detection of the coherent elastic scattering between a neutrino and the silicon nuclei. The array of low-noise fully depleted scientific CCDs was developed by the LBNL Micro Systems Labs with specific modifications that provide a better efficiency in particle physics detection. The thickness of the CCDs was increased to $675 \; \mathrm{\mu m}$ to amplify the active mass region of interaction, and the sensors are fully depleted with the help of the high-resistivity ($10 \text{k} \Omega$-m) silicon wafers.

On one hand, the differential event rate shown in Eq. (\ref{eq:diffEventRate}) is expressed as a function of the nuclear recoil energy $E_{nr}$. On the other hand, CCD sensors measure the ionization energy $E_{I}$, i.e the fraction of the silicon nuclei recoil energy that is converted into charge carriers. In order to relate the true recoil energy into the observable ionization energy we need to use the quenching factor $Q$ 
\begin{equation}\label{eq:quenching}
    Q = \frac{E_{I}}{E_{nr}}\;.
\end{equation}
In this way we can express the differential event rate in terms of the ionization energy
\begin{equation}\label{eq:diffEventRatequenching}
   \dv{R_0}{E_{I}} {(E_{I})}= \dv{R_0}{E_{nr}} \dv{E_{nr}}{E_I} = \dv{R_0}{E_{nr}} \frac{1}{Q} \Big( 1- \frac{E_I}{Q} \dv{Q}{E_{I}}\Big) \;.
\end{equation}
For this purpose we employed the quenching factor measurements by Chavarria \textit{et al.} \cite{Chavarria:2016xsi}, which can be fitted by 
\begin{equation}\label{eq:functionquenching}
   Q(E_{I}) = \frac{c_0 E_{I}+ c_1 E_{I}^2+ E_{I}^3}{c_2 + c_3 E_{I} + c_4 E_{I}^2}\;,
\end{equation}
where the parameters are $c_0 = 168\;  \text{keV}^2$, $c_1 = 156\;  \text{keV}$, $c_2 = 56 \;  \text{keV}^3$, $c_3 = 1097 \;  \text{keV}^2$ and $c_4 = 383\;  \text{keV}$ \cite{Aguilar-Arevalo:2019zme}.

Finally, for a realistic event rate calculation, one must take into account the reconstruction efficiency $\epsilon(E_{I})$ of the events registered in the CCDs:
\begin{equation}\label{eq:diffEventRateEff}
    \dv{R}{E_{I}} = \epsilon(E_{I}) \times \dv{R_0}{E_{I}} {(E_{I})} \; .
\end{equation}
For the CONNIE processing tools, $\epsilon(E_{I})$ has been evaluated in \cite{Aguilar-Arevalo:2019jlr} using simulated events, being fitted by 
\begin{equation}
    \epsilon(E_{I}) = b - [1+ e^{b_0 (E_{I}-b_1)}]^{-1} \; ,
\end{equation}
where $b=0.845$, $b_0  = 42.66$ and $b_1 = 0.067$.

The upper panel of Figure \ref{fig:eventRateNakWeak} shows the results of the differential event rate as a function of the ionization energy for a supernova located $10$ kpc away considering the Nakazato weakest model and $1$ kilogram of silicon detector mass with CONNIE efficiency. The distance of $10$ kpc corresponds approximately to the distance between the Earth and the center of the galaxy. The lower panel of Figure \ref{fig:eventRateNakWeak} shows the same, but considering the Livermore model and a SN at a distance of $196$ pc, which corresponds to the distance to the red supergiant Betelgeuse. From the figure, it can be seen that, although the slightly enhancement of $\nu_e$ flux at low energies, in general all flavors have similar outputs. However, when considering the contribution from all muonic and tauonic neutrinos, i.e. four times the $\nu_x$ curve, one can see that their signal is greater than the joined signal from $\nu_e$ and $\bar \nu_e$. 

The total number of observable SN neutrino events of all flavors $N_{obs}$ can be obtained by integrating the realistic differential event rate over all values of ionization energies:
\begin{equation}
    N_{obs} = \int_{E_{th}}^{\infty} \dv{R}{E_{I}}(E_{I}) \mathrm{d}E_{I} \; ,
\end{equation}
\noindent where $E_{th}$ is the threshold energy of the experiment under consideration. In the case of CONNIE, $E_{th} =  64 \;\text{eV}$.

\begin{table}[]
\bgroup
\def\arraystretch{1.2}
\begin{tabular}{cc|cccc|}
\cline{3-6}
\multicolumn{1}{l}{}                                     & \multicolumn{1}{l|}{}         & \multicolumn{4}{c|}{Supernova Model}                                                                                                                                                                                                                                               \\ \hline
\multicolumn{1}{|c}{\multirow{2}{*}{$M_{det}${[}kg{]}}} & \multirow{2}{*}{$d${[}pc{]}} & \multirow{2}{*}{Livermore} & \multirow{2}{*}{\begin{tabular}[c]{@{}c@{}}Nakazato \\ (weakest)\end{tabular}} & \multirow{2}{*}{\begin{tabular}[c]{@{}c@{}}Nakazato\\  (brightest)\end{tabular}} & \multirow{2}{*}{\begin{tabular}[c]{@{}c@{}}Nakazato\\  (black hole)\end{tabular}} \\
\multicolumn{1}{|c}{}                                    &                               &                            &                                                                                &                                                                                  &                                                                                   \\ \hline
\multicolumn{1}{|c}{1}                                   & $10^3$                        & 0.005                      & 0.001                                                                          & 0.002                                                                            & 0.005                                                                             \\
\multicolumn{1}{|c}{1}                                   & 196                           & 12.41                      & 3.13                                                                           & 6.17                                                                             & 12.49                                                                             \\
\multicolumn{1}{|c}{10}                                  & $10^3$                        & 0.048                      & 0.012                                                                          & 0.024                                                                            & 0.048                                                                             \\
\multicolumn{1}{|c}{10}                                  & 196                           & 124.08                     & 31.28                                                                         & 61.65                                                                            & 124.86                                                                            \\
\multicolumn{1}{|c}{30}                                  & $10^3$                        & 0.14                       & 0.04                                                                           & 0.07                                                                             & 0.14                                                                              \\
\multicolumn{1}{|c}{30}                                  & 196                           & 372.24                     & 93.83                                                                         & 184.95                                                                           & 374.58                                                                           \\ \hline
\end{tabular}
\egroup
\caption{\label{tab:results}
Number of observable supernova neutrino events, $N_{obs}$, calculated with CONNIE processing efficiency for three different values of silicon detector mass $M_{det}$, two supernova distances $d$ and considering four different SN models.} 
\end{table}

Table \ref{tab:results} summarizes the results for the number of observable events, $N_{obs}$, obtained with the four different SN models employed in this paper and considering different configurations of detector mass, $M_{det}$, and supernova distance, $d$. The CONNIE processing efficiency was also used. Due to the fact that $\nu_x (=\nu_{\mu} = \bar \nu_{\mu} = \nu_{\tau} = \bar \nu_{\tau})$, in the calculation of $N_{obs}$ the number of observable $\nu_e$ and $\bar \nu_e$ events was summed with four times the number of $\nu_x$ events.

It is important to highlight that the CONNIE experiment is a surface detector close to a nuclear power plant. This means that the background contamination is very high, reaching values such as 10 kdru (1 dru = 1 event/day/kg/keV). Therefore, considering the results of Table \ref{tab:results}, it would be very unlikely to distinguish events even for a very close SN.

Underground experiments are actually the best choice for SN$\nu$ detection. Indeed, there are several future prospects for CCD experiments with the purpose of direct detection of sub-GeV dark matter (DM). These experiments will be assemble in very low background environments with active mass going from $0.1$ kg to $10$ kg in different construction time scales.

The SENSEI (Sub-Electron-Noise Skipper-CCD Experimental Instrument) experiment is the first one to employ the Skipper-CCD technology, recently developed by Tiffenberg \textit{et al.} \cite{Tiffenberg_2017}. These improved CCDs are designed with a non-destructive readout system that allows for an ultra-low readout noise of $0.068$ $e^-$ rms/pixel, so that they have the capability of recording precisely the number of electrons in each pixel, increasing the sensitivity and allowing for the detection of low-energy neutrinos more effectively. A small prototype with a single Skipper-CCD was tested at Fermilab \cite{Crisler_2018}. Currently, a $100$ g Skipper-CCD experiment is being assembled at SNOLAB with a suppressed background rate of $5$ dru.

A larger mass experiment under construction is DAMIC-M (DArk Matter In CCDs at Modane) experiment, which will be assembled at the Laboratoire Souterrain de Modane (LSM), in France. The experiment will consist of a 1 kg detector composed of 50 CCDs with Skipper technology. The background contamination projection is of the order of 0.1 dru.

Finally, a major step in the sensitivity to sub-GeV DM interactions with electrons will be reached by the OSCURA project \footnote{The OSCURA project is an R$\&$D effort supported by Department of Energy to develop a 10 kg Skipper-CCD dark matter search experiment.}. It aims to build a $10$ kg Skipper-CCD experiment with a projection of $\sim 0.01$ dru background event rate.

\begin{table}[]
\bgroup
\def\arraystretch{1.2}
\begin{tabular}{cc|cccc|}
\cline{3-6}
\multicolumn{1}{l}{}                                     & \multicolumn{1}{l|}{}         & \multicolumn{4}{c|}{Supernova Model}                                                                                                                                                                                                                                               \\ \hline
\multicolumn{1}{|c}{\multirow{3}{*}{$M_{det}$ {[}kg{]}}} & \multirow{3}{*}{$d$ {[}pc{]}} & \multirow{3}{*}{Livermore} & \multirow{3}{*}{\begin{tabular}[c]{@{}c@{}}Nakazato \\ (weakest)\end{tabular}} & \multirow{3}{*}{\begin{tabular}[c]{@{}c@{}}Nakazato\\  (brightest)\end{tabular}} & \multirow{3}{*}{\begin{tabular}[c]{@{}c@{}}Nakazato\\  (black hole)\end{tabular}} \\
\multicolumn{1}{|c}{}                                    &                               &                            &                                                                                &                                                                                  &                                                                                   \\
\multicolumn{1}{|c}{}                                    &                               &                            &                                                                                &                                                                                  &                                                                                   \\ \hline
\multicolumn{1}{|c}{0.1}                                 & $10^3$                        & 0.0006                      & 0.0002                                                                         & 0.0003                                                                           & 0.0006                                                                             \\
\multicolumn{1}{|c}{0.1}                                 & 196                           & 1.50                       & 0.39                                                                           & 0.76                                                                             & 1.48                                                                              \\
\multicolumn{1}{|c}{1}                                   & $10^3$                        & 0.006                      & 0.002                                                                          & 0.003                                                                            & 0.006                                                                             \\
\multicolumn{1}{|c}{1}                                   & 196                           & 15.00                      & 3.92                                                                           & 7.62                                                                             & 14.85                                                                             \\
\multicolumn{1}{|c}{10}                                  & $10^3$                        & 0.06                       & 0.02                                                                           & 0.03                                                                             & 0.06                                                                              \\
\multicolumn{1}{|c}{10}                                  & 196                           & 150.02                     & 39.2                                                                           & 76.24                                                                            & 148.48                                                                            \\ \hline
\end{tabular}
\egroup
\caption{\label{tab:resultsDM}Number of observable supernova neutrino events, $N_{obs}$, considering the efficiency of Skipper-CCDs, two supernova distances $d$ and four different SN models. We show the results for the current silicon detector mass prospects of $0.1$ kg for SENSEI, $1$ kg for DAMIC-M and $10$ kg for OSCURA.}
\end{table}

In Table \ref{tab:resultsDM} we present the results for the number of observable SN$\nu$ events considering these three projects and employing an efficiency of $100$\% above a detection threshold of $15$ eV (typical Skipper-CCD efficiency value). From these results one can see that with the available mass of SENSEI it would be unlikely to detect SN$\nu$, but we already see a considerable chance of detection for DAMIC-M and OSCURA.

The results presented in Table \ref{tab:resultsDM} illustrate the amount of silicon one needs, combined with how close to the Earth the SN must occur, so that an effective detection of supernova neutrinos using silicon technology is attainable. Considering, for example, a CCD detector array with $10$ kg of silicon active mass, for a detection of an event the supernova must occur at most $2$ kpc away from the Earth, taking into account the Livermore model. For the Nakazato weakest model, this distance must be less than $1$ kpc. The results are, of course, model dependent, but the numbers displayed in Table \ref{tab:resultsDM} set the ballpark. It is worth mentioning that CCD sensors work by accumulating events in a certain time window and then reading the pixels individually. This readout process takes, in the best of the cases, few minutes, so that CCDs still have poor time resolution.
 
One can compare our results of Table \ref{tab:resultsDM} with the ones of the number of observable SN neutrino events coherently scattered off xenon nuclei in the XMASS experiment \cite{Abe_2017}. The XMASS experiment is a $832$ kg liquid xenon dark matter scintillation detector located at the Kamioka Observatory, in the Kamioka mine, $1000$ m underneath the top of Mt. Ikenoyama \cite{Abe_2013}. The light produced in the noble liquid due to scintillation is detected with the help of $642$ photomultiplier tubes inside a copper vessel surrounded by a $10$ m diameter and $11$ m high cylinder filled with ultra pure water.

The XMASS collaboration reported that, for a galactic SN located $10$ kpc away from the Earth, the number of SN neutrinos observable events would range between $3$ and $21$, depending on the SN model. For a SN located at the same distance as Betelgeuse, they estimated that $N_{obs} \sim 10^4$. Comparing the XMASS results with the results presented here, it is clear that the difference in the number of events for, say, $1$ kg of mass is in the order of $10^3$. Of course, this balance was already expected mainly due to the fact that the XMASS experiment has $832$ kg of xenon mass, while current experiments employing CCDs still do not have much active mass. Furthermore, the cross section of CE$\nu$NS with xenon nuclei is higher than with silicon, since the differential cross section of CE$\nu$NS, Eq. (\ref{eq:crossSection}), is enhanced with the number of neutrons in the target nuclei. 

 \begin{figure}
\vspace{3mm}
\includegraphics[scale= 0.44]{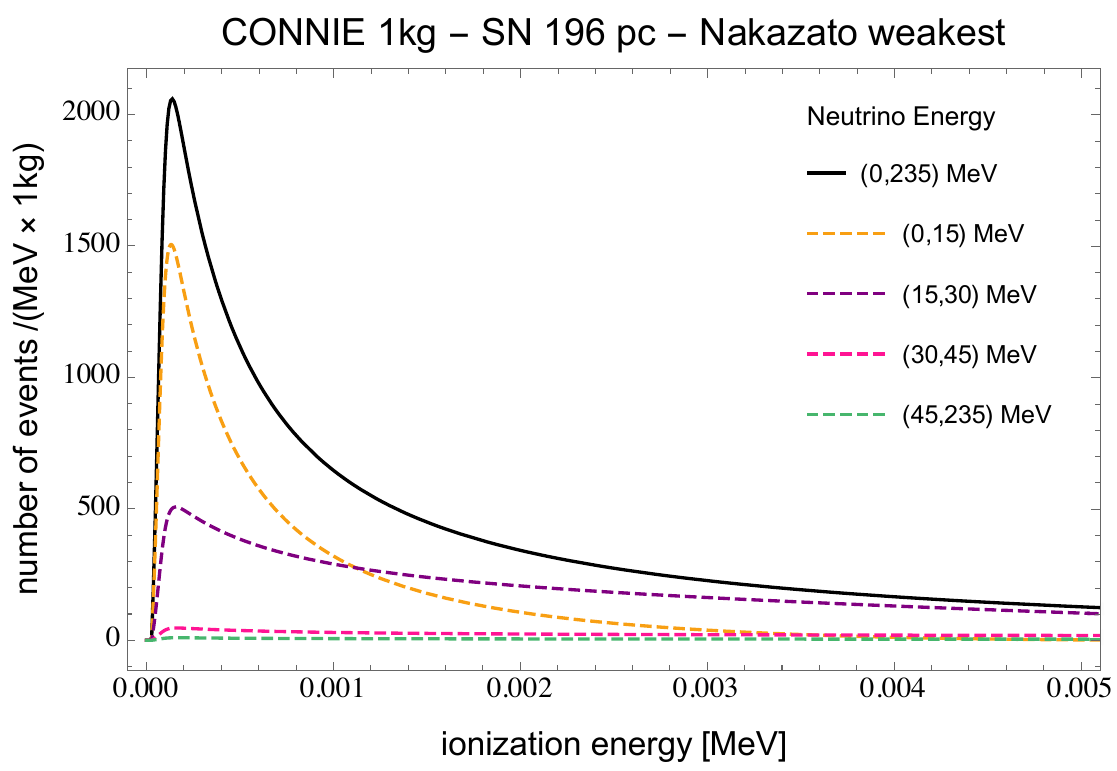}
\caption{\label{fig:eventRateNeutrinoEnergy}
Energy spectrum as a function of the silicon ionization energy considering different neutrino energy intervals. The curves show the contribution from all neutrino flavors to the event rate in that particular interval of $E_{\nu}$ and the CONNIE efficiency was also taken into account.}
\end{figure}

There are also other noble liquid experiments that would be sensitive to coherent scattering of SN$\nu$ \cite{Raj:2019sci}, such as the future experiments DARWIN \cite{Aalbers:2016jon} and ARGO \cite{Aalseth:2017fik}. Indeed, one of the first references to explore SN$\nu$ detection via this channel, using noble liquid target materials, was Ref. \cite{PhysRevD.68.023005}, where they considered the particular configuration of the formerly proposed CLEAN detector. The PICO collaboration also reported an estimate for CE$\nu$NS supernova neutrino detection considering the planned PICO-500 Bubble Chamber experiment \cite{Kozynets:2018dfo}. In addition, some of these large dark matter experiments have capability of pre-supernova neutrino detection \cite{Raj:2019wpy} (for more references in pre-SN$\nu$ detection, see e.g. \cite{Simpson_2019,Kato:2017ehj,Kato_2020}).

On the other hand, the benefit of CCDs compared to noble liquids is that a CCD detector array is sensitive to a $\lesssim 1$ keV threshold. Therefore, silicon can be used for detection of low energy SN neutrinos while the noble liquid is mostly sensitive to neutrinos above $15$ MeV. This fact is endorsed by Figure \ref{fig:eventRateNeutrinoEnergy}, that shows the contribution to the event rate of different neutrino energy intervals, considering all neutrino flavors and the CONNIE efficiency. The most relevant contribution comes from neutrinos with energies up to 15 MeV. Thus, silicon detectors are sensitive exactly in this low-energy regime in which other types of detectors lack efficiency. 

Besides, whereas water Cherenkov detectors and liquid scintillator detector experiments are primarily sensitive to $\bar \nu_e$ via the inverse beta decay (IBD) reaction, silicon detectors are sensitive to all flavors of neutrinos. Therefore, the latter can complement the information that would be recorded by present-day experiments, such as SuperKamiokande (SuperK) \cite{Ikeda_2007}, IceCube \cite{Abbasi:2011ss}, Borexino \cite{Cadonati_2002} and Baksan \cite{Domogatsky_2007}, or future experiments, such as JUNO \cite{An:2015jdp} and LENA \cite{LENA}.

\begin{figure}
\vspace{3mm}
\includegraphics[width=\linewidth]{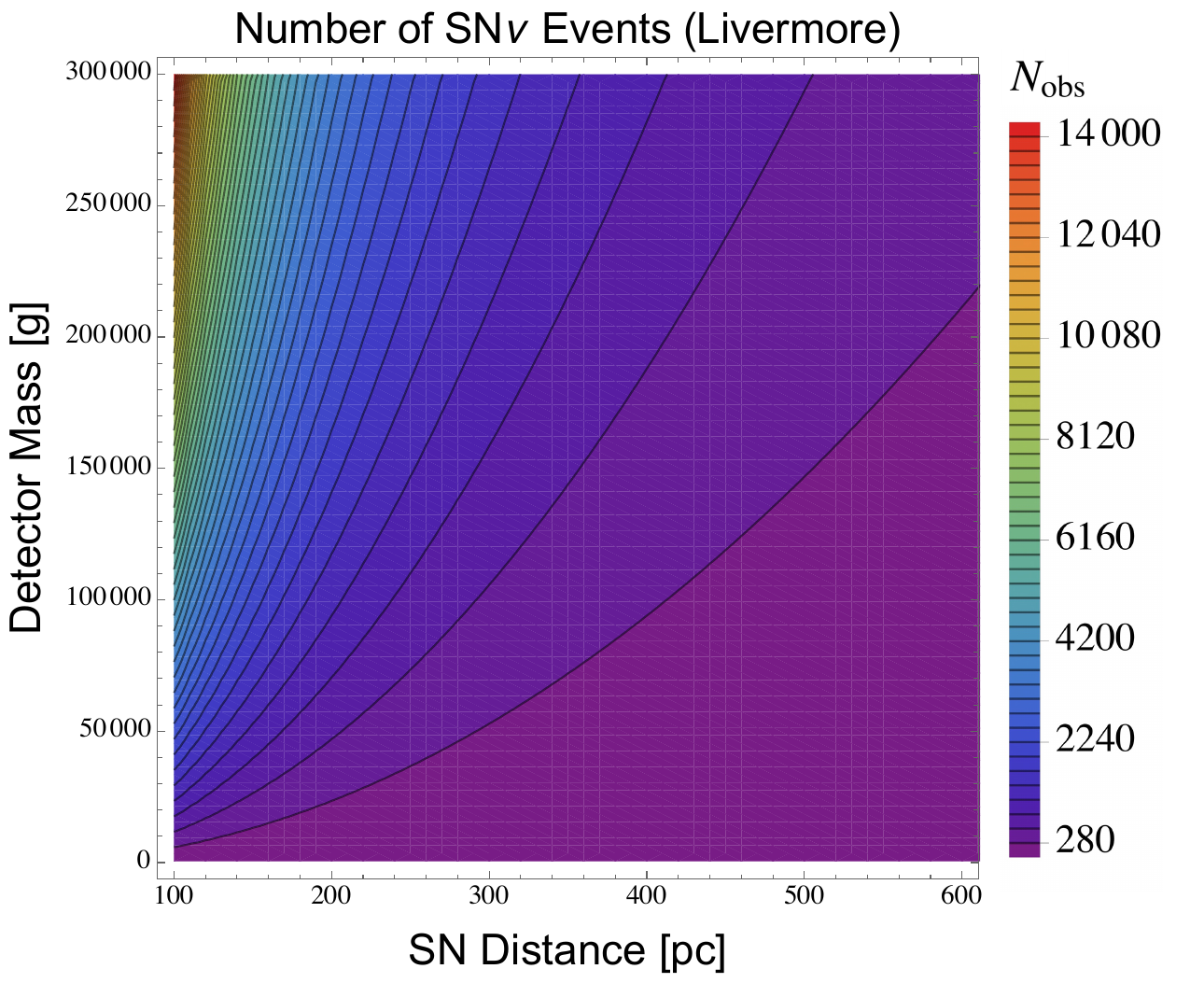}
\caption{\label{fig:NobsVariantLivermore} Contour plot of the number of observable supernova neutrino events, $N_{obs}$, for the Livermore model, and using CONNIE efficiency, when the detector mass $M_{det}$ is extrapolated to very large values.}
\end{figure}

Finally, we can compute the number of observable SN neutrino events, $N_{obs}$, coherently scattered off silicon nuclei considering a wide range of values for the detector mass and supernova distance. The results are displayed in a contour plot in Figure \ref{fig:NobsVariantLivermore}. In this case, the range of observable events shoots up to $10^4$ events, considering a very close SN and a very large detector mass. Of course, this amount of $M_{det}$ is still far beyond the current capabilities for CCD production, but we can not discard a great technological progress in the future.

\section{Outlook}
\label{outlook}

In the present work the number of observable supernova neutrino events detected via coherent scattering off silicon nuclei was estimated considering different mass values, SN distances and SN models. 

In the current scenario, the available experiments that employ silicon for the purpose of particle detection, mainly with the use of CCD sensors, do not have enough active mass for detection of neutrinos coming from SN further than $1$ kpc yet. 

However, a few experiments that use scientific CCDs for the purpose of particle detection will start operating in the near future, e.g. SENSEI and DAMIC-M. Both will have favorable conditions for SN$\nu$ detection and will employ the recent Skipper technology.

Furthermore, there is a great R$\&$D effort to build a 10 kg Skipper-CCD experiment (OSCURA) within the next years with the goal of sub-GeV dark matter detection. As discussed above, it will also have a good potential for supernova neutrino detection.

The main advantage of SN$\nu$ detection with CCDs, compared to other particle detectors, is the fact that CCD technology is sensitive in the very low-energy regime. Namely, CCDs can detect incident neutrinos with energies below 15 MeV, whereas other SN$\nu$ detection facilities will mostly detect higher-energy neutrinos.

In summary, with the currently available detectors, the detection of SN neutrinos coming from distances of the order of kpc is very unlikely and would have to wait for future upgrades to increase the active mass of the available CCD arrays. For a supernova located at a distance similar to that of Betelgeuse, the results show that, with the detector mass available in current and near future experiments, it would be possible to detect a considerable amount of SN neutrinos. This detection could provide crucial information on the dynamics of the supernova explosion mechanism, as well as bring insight into the process of formation and evolution of compact stars.

\begin{acknowledgments}
We would like to thank the CONNIE Collaboration, in special Juan Estrada, for useful discussions and comments. This work was partially supported by CAPES (Finance Code 001), CNPq, FAPERJ, and INCT-FNA (Process no. 464898/2014-5).

\end{acknowledgments}

\bibliography{SNneutrinos}

\end{document}